# Institute for Molecular Physics at the University of Maryland

Jan V. Sengers

Institute for Physical Science and Technology, University of Maryland, College Park, MD 20742

13 December 2013

1. Introduction

The Institute for Physical Science and Technology at the University of Maryland was founded in 1976 from a merger of the Institute for Fluid Dynamics and Applied Mathematics (IFDAM) and the Institute for Molecular Physics (IMP), which were established at the College Park Campus after World War II to enhance the expertise of the University of Maryland in some areas of science and technology of interest to the US Department of Defense. Here I try to reconstruct the history of the Institute for Molecular Physics at the University of Maryland. This account is based on documentation I have been able to collect during my tenure as a faculty member at the University of Maryland, supplemented with personal memories. In addition I have benefitted from discussions with a number of colleagues.

The establishment of IMP at the University of Maryland was motivated by a desire of the US Navy to replicate in the US the experimental facilities of the Van der Waals Laboratory at the University of Amsterdam. Hence, to understand the origin of IMP at the University of Maryland we need to start with a discussion of molecular physics at the University of Amsterdam.

2. Van der Waals Laboratory at the University of Amsterdam

The tradition of molecular physics at the University of Amsterdam was a heritage of Johannes Diderik van der Waals (1837-1923) [1]. Van der Waals received his doctorate in 1873 at the University in Leiden with a thesis *On the continuity of the gaseous and liquid states* in which he developed his well-known equation of state [2]. When the "Athenaeum Illustre" in Amsterdam was converted into a real university in 1877, Van der Waals became the first Professor of Physics at the University of Amsterdam [3]. The work of Van der Waals established a framework for understanding the transformation of fluids from the gaseous to the liquid state for one-component, binary and ternary mixtures including critical behavior [1,4]. His formulation of the principle of corresponding states enabled Heike Kamerlingh Onnes in Leiden to liquefy helium [4]. Van der Waals also developed the squared-gradient theory for fluid interfaces [5]. Van der Waals received the Nobel Prize in 1910 (Fig. 1).

Van der Waals, being a theoretician himself, saw also the need of experimental research in molecular science in general and at high pressures specifically. For this purpose a Van der Waals Fund was established in 1898 solicited from donations [7,8]. In 1908 Van der Waals was succeeded by his son Johannes Diderik van der Waals, Jr. for theoretical physics and by Philip A. Kohnstamm for applied experimental thermodynamics [3]. With the help of Kohnstamm, who took notes of the lectures of Van der Waals, Sr., two famous textbooks on the thermodynamics of fluids were published, one dealing with single-component fluids [9] and one dealing with mixtures [10]. Kohnstamm did start an experimental high-pressure research program in the existing Physical Laboratory of the University of Amsterdam in collaboration with the Belgian scientist J. Timmermans [11]. However, the interest of Kohnstamm drifted towards education and he became from 1919 more and more active in education in Amsterdam and in Utrecht [3]. Instead, it was A.M.J.F. Michels (1889-1969), a graduate student of Kohnstamm, who was destined to develop the initiatives of Van der Waals and Kohnstamm into a Van der Waals Laboratory with an international reputation [7,8].

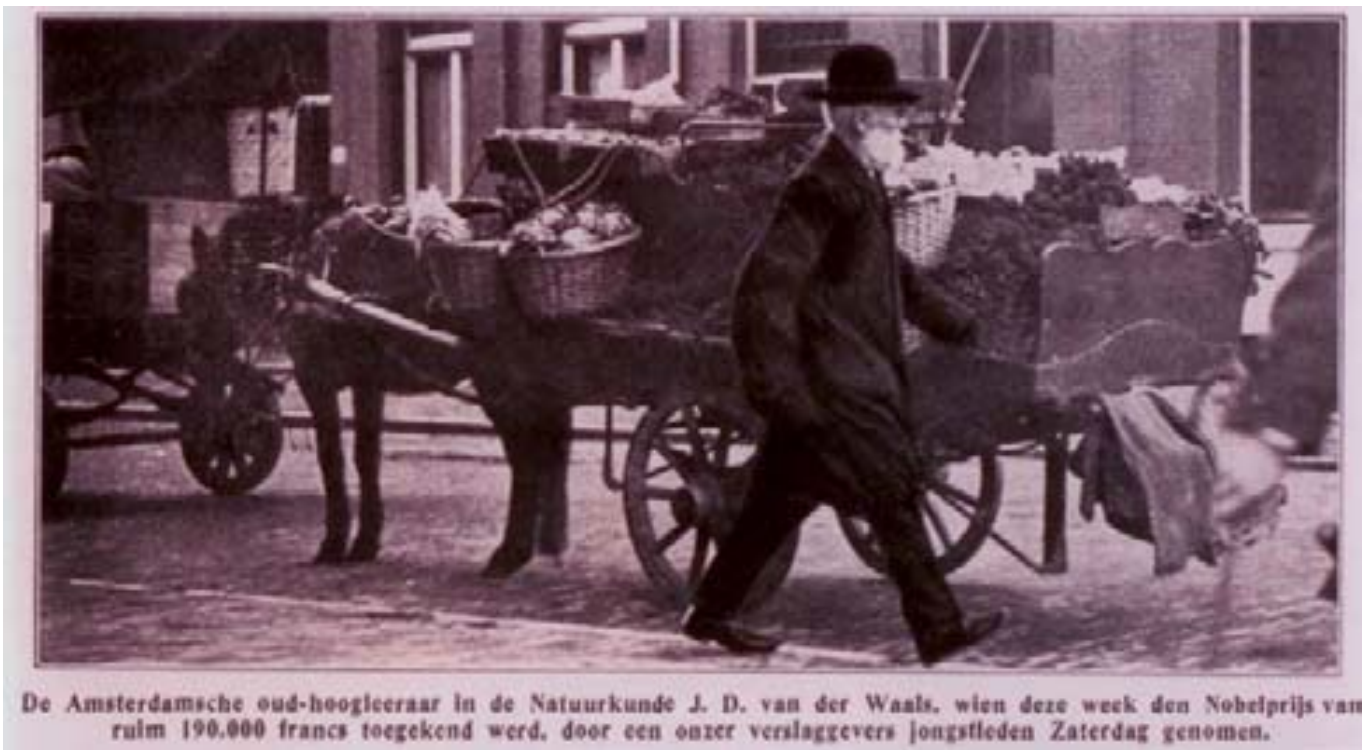

Figure 1: J.D. Van der Waals walking in Amsterdam just after having received the Nobel Prize [6] [News item in "Het Leven", April 10, 1910].

Antonius (Teun) Michels had a very strong personality with an extraordinary energy and drive to reach his objectives [12]. Starting as a Research Assistant of Kohnstamm in the Physical Laboratory of the University of Amsterdam, Michels embarked on an aggressive high-pressure research program. He had also remarkable engineering talents in a life-long companionship with J. Ph. Wassenaar, an instrument maker trained by Kamerlingh Onnes, which enabled him to develop new high-pressure equipment including the so-called Michels dead-weight gauge for accurate pressure measurements up to 3000 bar [7]. He received his Ph.D. in 1924 on the accurate measurements of pressure-volume-temperature isotherms.

Michels admired Kamerlingh Onnes. What Kamerlingh Onnes had meant for low-temperature physics, is what Michels wanted to become for high-pressure science. The financial resources of the Van der Waals Fund were limited, but this did not deter Michels who decided to raise additional funds himself. The research capabilities established by Michels caught the attention of Imperial Chemical Industries (ICI) in the United Kingdom and Michels began to receive research assignments from the chemical industry. ICI was hesitant giving money to the University of Amsterdam. Michels handled this by establishing an independent foundation, called "Boyle Stichting" for the receipt of financial support from industry [7].The interaction of Michels with ICI, especially with R.O. Gibson at ICI, led to the discovery of the polymerization of ethylene [13-16]. As a result of his success, Michels was appointed a lecturer at the University of Amsterdam in 1929 and a start was made building a new laboratory, separate from the Physical Laboratory, which was completed as the Van der Waals Laboratory in 1935 (Fig. 2). In 1939 Michels received an appointment as extraordinary professor in experimental and technical physics. This was a part-time position which enabled Michels to pursue additional activities, including eventually a part-time affiliation with the University of Maryland after World War II. In spite of some difficulties due to the Nazi occupation during World War II, the research at the Van der Waals Laboratory continued to flourish.

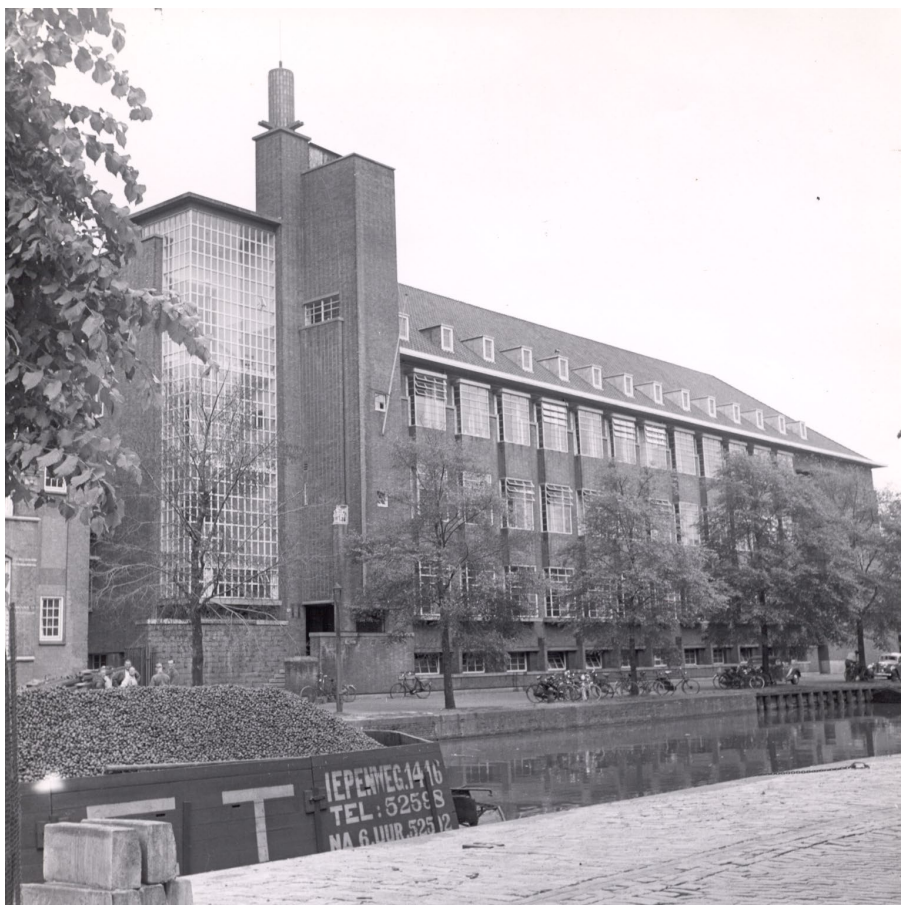

Figure 2. Van der Waals Laboratory located at the Nieuwe Achtergracht in Amsterdam (1935-1965) [Photo archives of Michels family].

After World War II, Michels established contact with a company "t Hart" in Rotterdam to market his high-pressure equipment (Figs. 3,4). For the interpretation of the experiments, Michels made sure that the Laboratory always had a competent theorist as a staff member. Scientists who have served in this capacity include Rudy J. Lunbeck, who became a professor at the Technical University Eindhoven, John A.M. Cox, who became a professor at Leiden University, and Martinus (Tiny) J. G. Veltman, who became a professor at Utrecht University and eventually received a Nobel Prize. In addition, Michels sought advice of theoreticians on a regular basis, such as Jan de Boer, Sybren de Groot and Peter Mazur, well-known scholars in statistical physics and thermodynamics. By the middle of the 20$^{th}$ century, the Van der Waals Laboratory had established an international reputation as a highly reliable source of thermophysical-property information for fluids as a function of pressure and temperature [7,8].

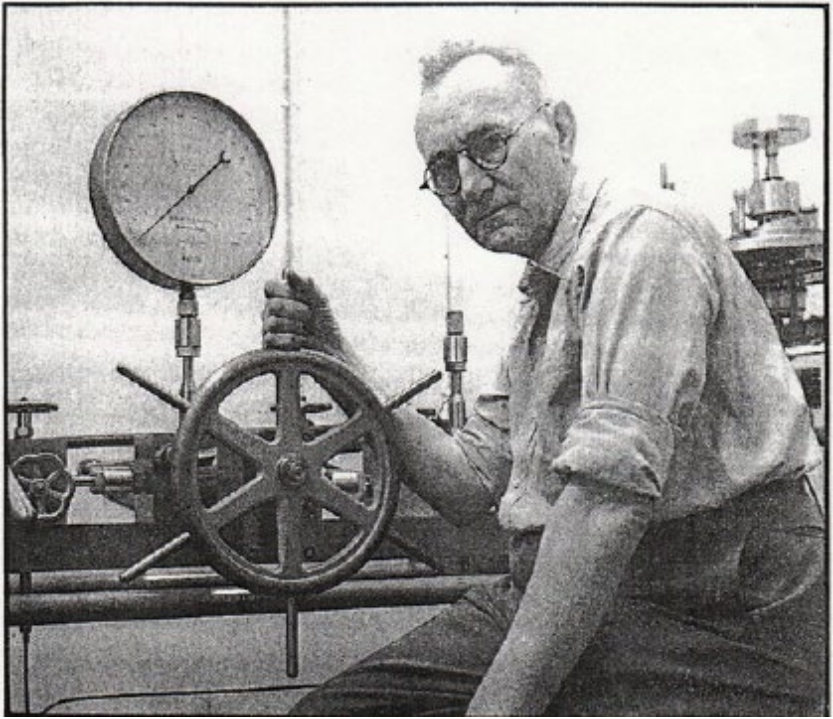

Figure 3. A.M.J.F. Michels in the Van der Waals Laboratory with his oil press for generating pressures up to a few thousand bar. In the far right is Michels' dead-weight gauge, shown in more detail in Fig. 4.

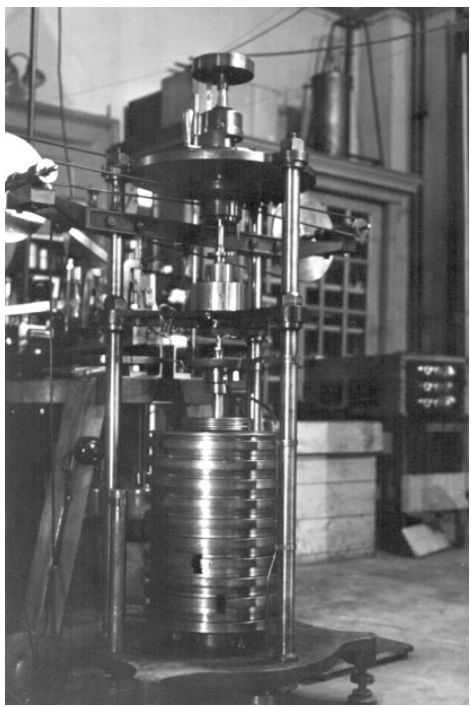

Figure 4. Michels' dead-weight gauge for measuring pressures up to a few thousand bar [Photo archives of Michels family].

3.  **Michels and the Institute for Molecular Physics**

In 1939 Michels also received an appointment as a reserve Major *Special Services* to pursue weapons related research for the Dutch Defense Department. During the occupation of the Netherlands by the Germans, Michels became very active in the underground resistance under the pseudonym "Thijs Smit" and became responsible for the resistance in the eastern part of Amsterdam. After narrowly escaping arrest in 1944, he went into hiding in Amsterdam. To excuse his absence from the University, his wife pretended that her husband was hospitalized after one of his customary visits to DSM (Dutch State Mines) in the southern part of the Netherlands which just had been liberated from the German occupation [12]. After the northern part of the Netherlands was also liberated in 1945, Michels was promoted to Colonel *Special Services* and became Head of a Military Mission of the Dutch Department of Economics to review German technology and science. This also led to regular visits of Michels to the US. As an undergraduate student I took two courses taught by Michels in Amsterdam. The courses would end early because Michels had to leave again for the US; the courses were completed by a staff member, Rudy J. Lunbeck, who actually was even a better teacher than Michels. For instance, Michels would cling to the thermodynamic notation of Van der Waals, like using the Greek symbols $\eta$ for entropy and $\varepsilon$ for energy. When Michels was off to the US, Rudy Lunbeck would tell us that we really should use the IUPAC notation, like *S* for entropy and *U* for energy.

As a result of the increasing visibility of Michels in the US, the Navy became interested in the research pursued at the Van der Waals Laboratory. During his assignment as Head of the Military Mission to Germany from 1945 till 1947, Michels had already befriended Rear Admiral Walter G. Schindler in Frankfurt who subsequently served as a commanding officer at the Naval Ordnance Laboratory (NOL) in White Oak, MD in the 1950's.  Michels also established contact with Thomas Killian, Science Director at the Office of Naval Research (ONR). ONR had a mandate to stimulate university-based research in science [17]. Tom Killian promised Michels that he would stimulate high-pressure research in the US and that he would send some scientists, each for a year, to interact with Michels at the Van der Waals Laboratory. The first one was Lt. Commander Thomas B. Owen, who was stationed administratively at the Office of the Naval Attaché in Amsterdam in 1950. With his expertise in chemistry and chemical engineering, a major assignment of Tom Owen was to keep abreast of the type of high-pressure work done at the Van der Waals Laboratory. Tom Owen was a war hero who had received the Silver Star for destroying cryptographic codes after his ship, the Cruiser of Northampton, was torpedoed off Savo Island in 1942. He also had survived the 1945 sinking of the destroyer Bush by Japanese kamikaze planes. Subsequently, as a Navy Rear Admiral, Tom Owen became Director of the Naval Research Laboratory (NRL) and later Chief of Naval Research. In Amsterdam Tom Owen was succeeded by Lt. Commander William (Bill) Ennis. In 1951 the London Office of Naval Research issued a technical report describing the research at high pressures at the Van der Waals Laboratory: Absorption spectra at high pressures, measurements of thermodynamic properties of argon, heat conductivity of gases, piezo-electric resonance in quartz, resonance shift in copper, low-temperature equation of state data, measurements of ferroelectrics and semiconductors, acoustic velocities, and thermodynamic measurements of mixtures [18]. This report was distributed to all major science and technical agencies of the US Government.

As one of the projects at the Van der Waals Laboratory, Michels had developed an apparatus for studying the rapid expansion of compressed gases behind a piston, a subject of great technical interest to NOL.  Hence, NOL sent several members of its scientific staff to the Van der Waals Laboratory in Amsterdam, namely, Paul A. Thurston, Arnold E. Seigel, Sigmund J. Jacobs, and Zaka I. Slawsky [19-21]. Seigel [22] and Jacobs [23] even received a Ph.D. at the University of Amsterdam (Fig. 5). In The Netherlands, the new Ph.D. customarily hosts an elegant dinner with Ph.D. advisor, colleagues, family and friends. The guests at the Ph.D. dinner of Arnold Seigel included Jan Burgers, Professor at the Technical University Delft, who later would become a professor in IFDAM at the University of Maryland.

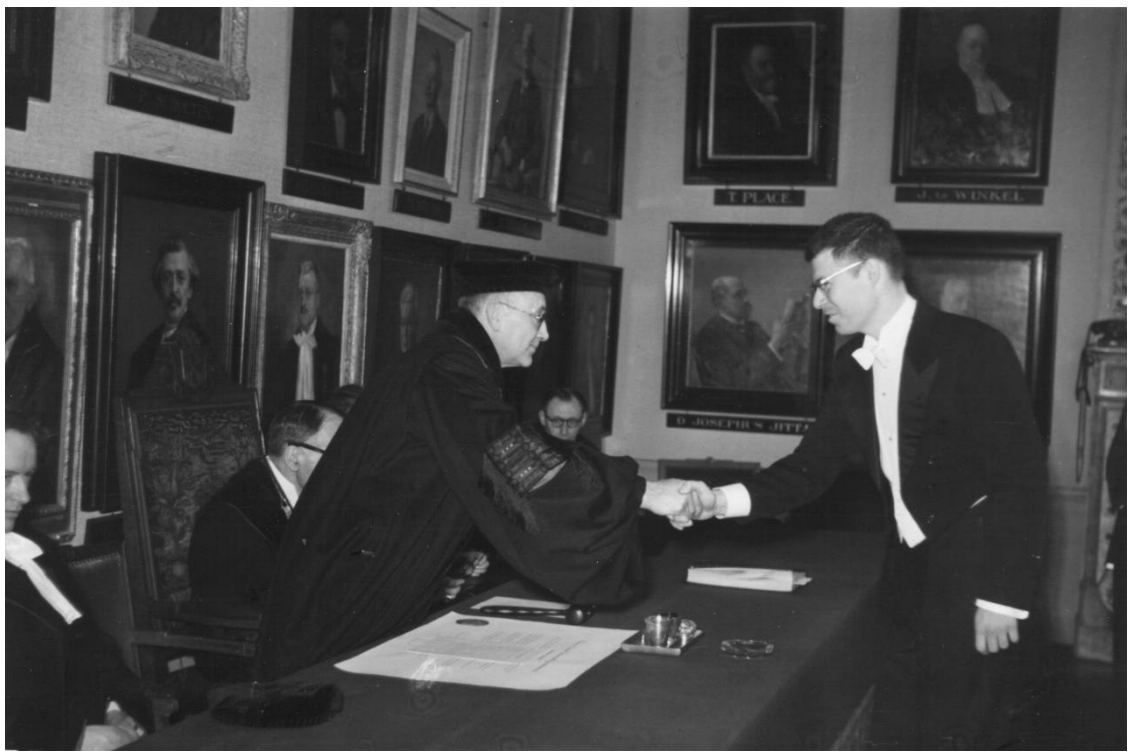

Figure 5. Arnold Seigel of NOL (right) being congratulated by Michels (left) upon receipt of his Ph.D. at the University of Amsterdam [Photo archives of Michels family].

Michels had also established a connection with the chemical industry in the US and at about the same time, Jack M. Lupton of DuPont received a Ph.D. at the Van der Waals Laboratory [24]. In addition, another person from the US, Abraham S. Friedman, spent a year (1951/1952) at the Van der Waals Laboratory before starting a staff position in the Heat Division of the National Bureau of Standards (NBS) in Washington, DC. The author, Jan V. Sengers, was a graduate student at the Van der Waals Laboratory from 1952 till 1962 (Fig. 6). When I joined the Van der Waals Laboratory in 1952, my first assignment was to complete the work of Abe Friedman [25]. Many scientists from the United States would visit the Van der Waals Laboratory in the 1950's, including William L. Marshall from the Oak Ridge National Laboratory as a Guggenheim Fellow in 1956-1957.

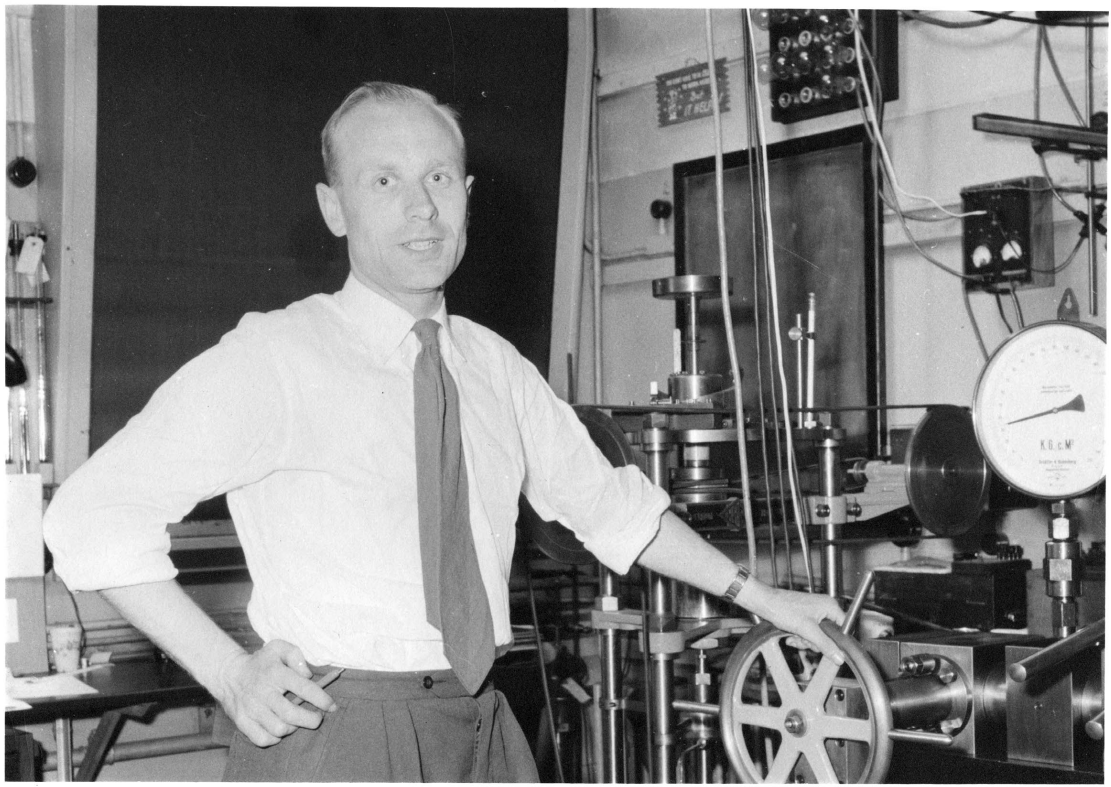

Figure 6. The author, Jan V. Sengers, as a graduate student in the Van der Waals laboratory
[AIP Emilio Segrè Visual Archives, Sengers Collection].

The political future in Western Europe was somewhat uncertain after World War II. Communist parties were active in a number of countries with a majority in France and Italy. The thought arose that a facility similar to the Van der Waals Laboratory should be available in the US in case the Soviet Union would invade Western Europe. Knowing Michels, I would not be surprised if this idea did not originate with NOL, but with Michels himself. In any case, Zaka Slawsky of NOL was assigned the task of preparing a proposal to ONR which he did together with Arnold Seigel. And indeed, in 1952, an agreement was reached between representatives of ONR and the University of Amsterdam that the latter (Michels, of course) would duplicate the Van der Waals Laboratory at the University of Maryland in College Park, an ideal location because of its proximity to NOL in White Oak, MD. The rational for the decision was stated as: *Over a period of years the laboratory in Amsterdam has developed special techniques for the study of molecular interactions, in particular methods for precise measurements at high pressures. In view of the importance of this work to science generally and to Ordnance in particular, the facilities in Amsterdam for this work should be available in this country in the event Holland was ever overrun by a hostile power.* Financial support for the new laboratory would be shared by ONR and by the University of Maryland. To implement the agreement, Michels was appointed Professor at the University of Maryland, while also keeping his position as Extraordinary Professor at the University of Amsterdam. At the groundbreaking of the new facility the first shovel of dirt was dug by J.D. van der Waals, Jr. (Fig. 7).

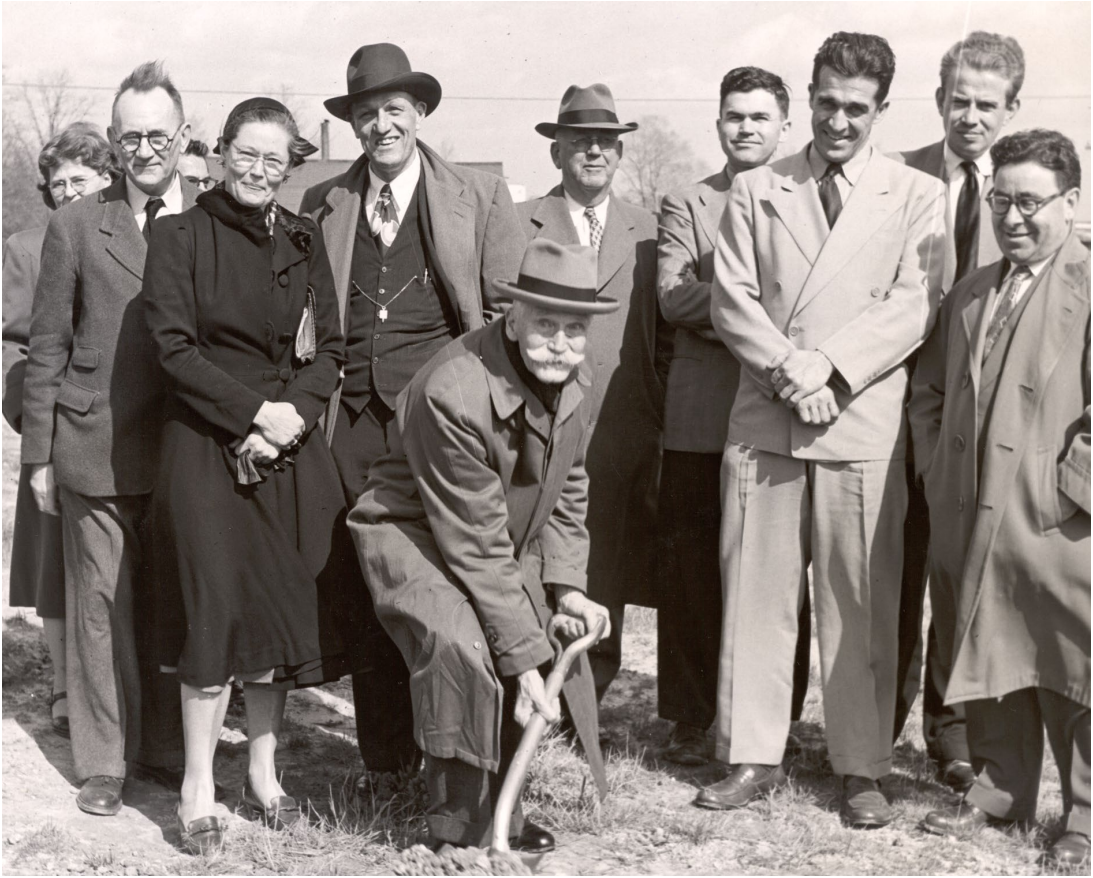

Figure 7. Ground breaking of the Institute for Molecular Physics by J.D. van der Waals, Jr. in 1953. From left to right: #2 A.M.J.F. Michels, #3 a daughter of J.D. van der Waals, Jr., #4 Nathan L. Drake, Head of the Chemistry Department, #5 J.D. van der Waals, Jr. From right to left: #1 Zaka I. Slawsky, #2 Laurens Jansen, #3 Berger M. Shepard, #4 Paul A. Thurston [Photo archives of Michels family].

The laboratories in the basement got vibration-free central pads with foundations separate from those of the building itself. To facilitate accurate temperature control, none of the laboratories would have windows. This feature caused the outside appearance of the building to be less elegant than that of the original Van der Waals Laboratory (Fig.2), but made the laboratories also ideal for optical research. The building, housing the Institute for Molecular Physics, was completed in 1954 (Figs. 8, 9). In 1976 its name was changed to IPST building.

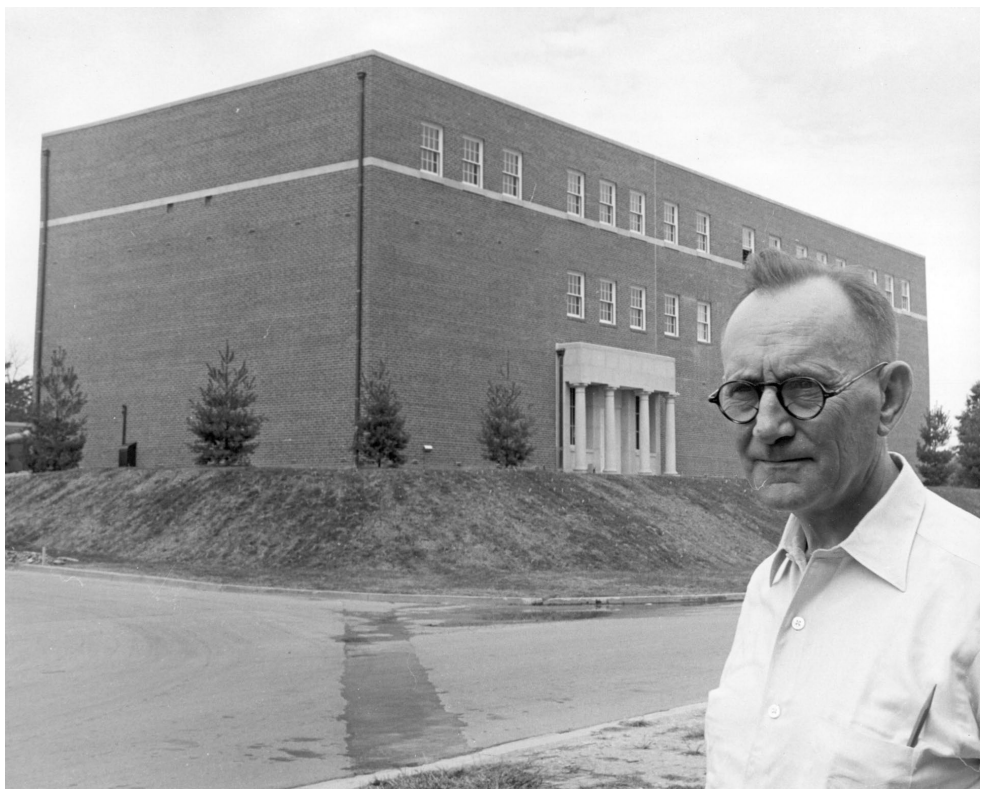

Figure 8. Michels and the new Institute for Molecular Physics [Photo archives of Michels family].

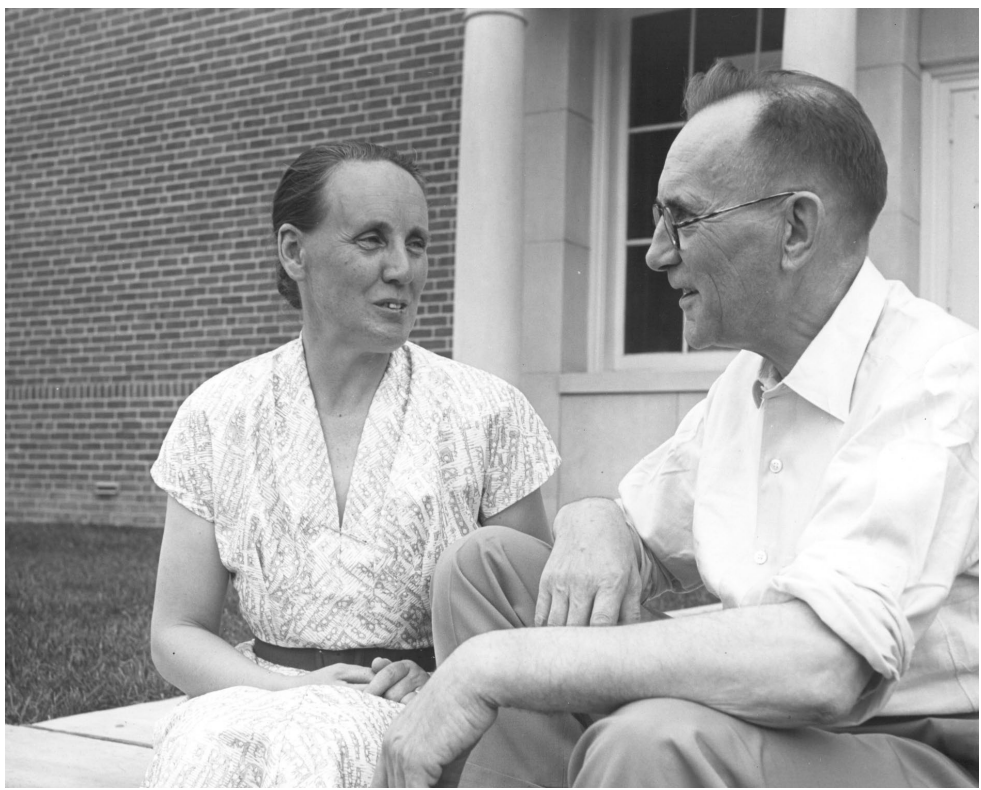

Figure 9. Professor A.M.J.F. Michels and his wife, Dr. C.A.M. Michels-Veraart, in front of the new Institute for Molecular Physics [8] [Photo archives of Michels family].

The new Institute for Molecular Physics was administratively located in the Department of Chemistry of the University of Maryland with the Head of Chemistry, Nathan L. Drake, also serving as the Administrative Director of the Institute. At the same time Michels, as Professor at the University of Maryland, was given a mandate to design the new Institute and to develop its research program. High-pressure equipment produced by 't Hart in Rotterdam, a joint venture with Michels , was shipped to the University of Maryland. The mass of steel plates with weights up to 25 kg, to be loaded onto Michels' dead-weight gauges as shown in Fig. 4, needed to be measured in the laboratory with a precision of 1:100,000. A mass balance, purchased for this purpose in Switzerland, arrived one day at the Van der Waals laboratory in Amsterdam. As a graduate student I was told to check and calibrate this mass balance before it would be shipped to the University of Maryland. This mass balance is now a museum piece located in the IPST building at the University of Maryland.

Arnold Seigel and Zaka Slawsky from NOL became affiliated with the Institute for Molecular Physics as consultants, while Berger M. Shepard in Zaka's Ballistic Department at NOL, provided technical support for the high-pressure equipment. An additional full-time technician, Howard J. Kopp, was hired in 1954. Mr. Kopp started his career at the Institute for Molecular Physics with a visit to the Van der Waals Laboratory in Amsterdam. A glass blower, John W. Trembly, was hired in 1955.

The first experimental scientist appointed in the new institute was Homer W. Schamp, Jr. During World War II Schamp had been drafted in the Navy and was assigned to work at NOL with Zaka Slawsky. After the war had ended, Zaka Slawsky recommended Schamp to the Chair of Physics at the University of Michigan to pursue a Ph.D. degree in physics. After his graduation, Zaka Slawsky brought Schamp in contact with Michels in the hope that Schamp would come to the new Institute for Molecular Physics. Although Schamp did not consider Michels to be a sympathetic person, Slawsky was able to induce him to come to the Institute at the University of Maryland anyway. As a first step, Schamp spent some time at the van der Waals laboratory in Amsterdam to become familiar with measuring pressure-volume-temperature isotherms at high pressures [26].

For the theoretical support of the research Michels recruited Laurens Jansen, a graduate student of Sybren de Groot in Leiden [27]. The first series of papers involved a triangle involving the Institute for Molecular Physics (Jansen), the Van der Waals Laboratory (Michels) and DuPont (Lupton) [28]. During his appointment at the Institute, Jansen also wrote his Ph.D. thesis on *Some Aspects of Molecular Interactions in Dense Media* which was completed in 1955. In 1955 a second theorist was appointed, namely Edward A. Mason. Mason had received a Ph.D. with Isadore Amdur at MIT and had been a postdoc with Joe Hirschfelder at the University of Wisconsin [29].

Because of his two positions, one at the University of Amsterdam and one at the University of Maryland, Michels  traveled  from Amsterdam to College Park several times each year. This type of commuting between Europe and the US was rather unusual in those years. Not surprisingly, therefore, a (US) newspaper article, written by June Grant, appeared under the title *The Flying Dutchman in Person: This One Flies 25,000 Miles Yearly Between Jobs in Holland and Maryland*.

Michels had also been Chair of the first Gordon Conference on High-Pressure Research held in Meriden, NH in 1955. Invited speakers included Robert W. Zwanzig, at that time at The Johns Hopkins University, and Elliott W. Montroll from the University of Maryland. However, Michels had failed to invite Percy W. Bridgman, a most prominent high-pressure researcher in the US. This was unacceptable and E.W. Comings of Purdue University was asked to organize a second Gordon Conference on High-Pressure Research to be held in 1956. Comings invited both Bridgman and Michels to lecture at this Gordon Conference. Michels, instead of going himself, sent by me to this Gordon Conference to give two presentations. Unlike Michels, I had to cross the Atlantic by boat, since that was a cheaper mode of transportation at the time. One problem was that European-size 5 cm slides were not compatible with American slides of 3.5 inches. Michels had solved this problem by making sure that of each slide in the Van der Waals Laboratory an American-size copy was kept in the Institute for Molecular Physics at the University of Maryland. Hence, after arriving in New York, I first had to travel to College Park, MD to get the slides for my presentations at the Gordon Conference in New Hampshire. At my visit to the Institute in 1956, I not only met again Homer Schamp and Mr. Kopp, known to me from their earlier visits to the Van der Waals Laboratory, but I also became acquainted with a graduate student, Barbara Castle, who shortly thereafter spent a year at the Van der Waals Laboratory in Amsterdam [30] as part of the interaction between the two institutions. I also used this opportunity to visit Bob Zwanzig at the Johns Hopkins University in Baltimore: later Bob would become a professor at the University of Maryland. At the Gordon Conference I saw some American luminaries in high-pressure research, like Bridgman and Harry G. Drickamer. I also met Jack Lupton from DuPont again; upon my return in Amsterdam I was told that Lupton had sent some favorable comments about my presentations at the Gordon Conference to Michels. Bob Zwanzig was also at this second Gordon Conference on High-Pressure Research. Another participant of interest to me, was Jan C. Stryland, who had been a deputy director under Michels at the van der Waals Laboratory for many years, but had become a professor at the University of Toronto. After the conference, I traveled with Stryland to Toronto before returning to Amsterdam.

**4. Review of the Institute for Molecular Physics in 1956**

In 1956 a crisis had developed concerning the Institute for Molecular Physics. Michels and Drake did not get along very well. With Drake, Head of the Chemistry Department, as the Administrative Director, there was no clear agreement between Michels and Drake on the final authority, which caused tensions about the management of the Institute. The Institute had been supported by ONR with research contracts for the original contracting period from April 1, 1953 till December 31, 1955. After the expiration of this contract period, ONR felt that with the advent of new weapons its original interest in work at high pressures had diminished and that the national interest indicated the need for a substantial reduction of its financial support. In response to this problem the University of Maryland established a prestigious Ad Hoc Committee to review the Institute for Molecular Physics. The members of this Ad Hoc Committee are listed in Table 1.

Table 1. Members of the Ad Hoc Committee to review the Institute for Molecular Physics

Dean Ronald Bamford, Graduate School, University of Maryland

F.G. Brickwedde, Department of Physics, University of Maryland

M.H. Martin, Director of IFDAM at the University of Maryland and Committee Chair

Dean Leon P. Smith, College of Arts and Sciences, University of Maryland

C.E. Sunderlin, Deputy Director, National Science Foundation, Washington DC

Specifically, the Committee considered the following two questions: *What is the proper position of the Institute for Molecular Physics in the University of Maryland structure* and: *How should the work in the Institute for Molecular Physics be administered to attain the maximum possible effect?* [31]. The Committee observed that *The Institute has a fine building, excellent equipment, and the nucleus of a good staff. A number of credible research papers has come from the Institute, in spite of its infancy, and its senior staff has participated in graduate teaching in both the Physics and Chemistry departments*. The general recommendations of the Committee were as follows:

1. The Institute for Molecular Physics should be placed under a full-time Director.
2. An ad hoc Advisory Panel for Molecular Physics should be appointed to advise the President on nominations for the post of Director.
3. The Institute for Molecular Physics should be a separate Institute in the University structure, with the Director reporting directly to the Dean of the College of Arts and Sciences or to the Dean of the Graduate School.
4. Pending action by the Advisory Panel, the Institute for Molecular Physics should remain under the direction of N.L. Drake (after consultation with the staff of the Institute).
5. The original objective of the Institute, namely, that the "know how, the philosophy and approach to research" of the Van der Waals Laboratory "should be taken over and followed" should be enlarged to permit the free reign of scholarly inquiry within the legitimate interests of the staff comprising the Institute.

In addition, the Committee made some specific recommendations in support of the objectives listed above. Michels responded to an invitation to provide his account of the history of the Institute for Molecular Physics and his thoughts for the future of the Institute [32]. The committee recommended that *Michels should be urged to continue as Research Professor part time*, because of his valuable contributions to the research program of the Institute. However, Michels accepted a full-time professorship position at the University of Amsterdam instead [12]. Shortly thereafter, Seigel, Slawsky, and Shepard of NOL ceased their affiliation with the Institute. In addition to his research position at NOL, Seigel became a part-time lecturer in the Department of Mechanical Engineering of the University of Maryland. After his retirement from his position of Chief of the Ballistic Department at NOL, Seigel became a visiting professor at the University of Maryland and in 1978 took charge as Director of the Instructional Television System at the University of Maryland. After their retirement form NOL in 1975, Zaka Slawsky and his twin brother, Milton

Slawsky, established the Mollie and Simon Slawsky Memorial Tutoring Clinic in the Department of Physics at the University of Maryland (Fig. 10). An interesting detail is that the spouses of the twin brothers Slawsky were sisters.

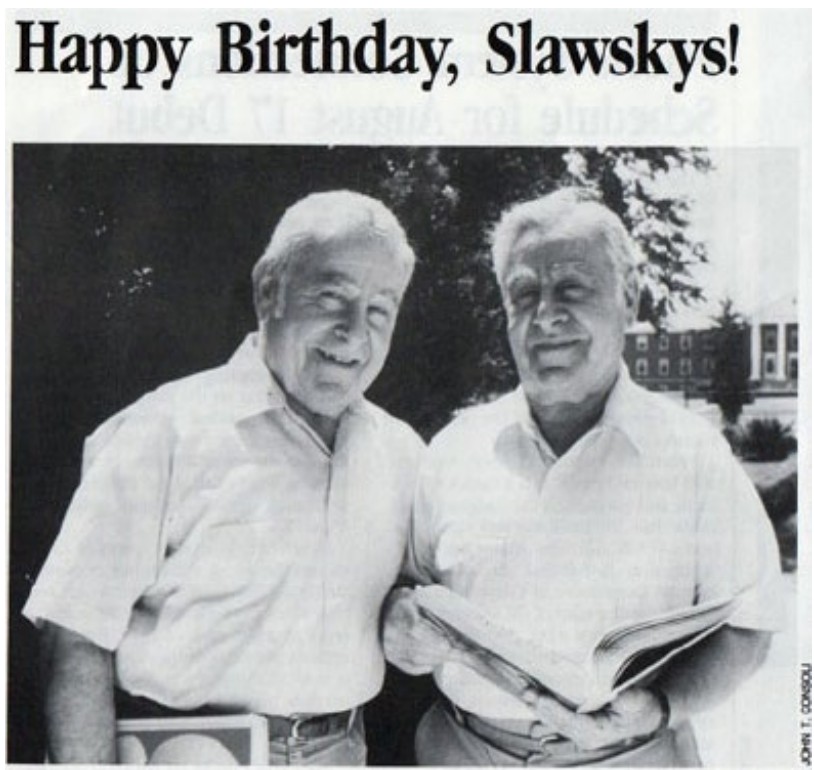

Figure 10. The twin brothers, Zaka and Milton Slawsky, celebrating their 80$^{th}$ birthday in 1990 at the University of Maryland.

6. **The Institute for Molecular Physics: 1956-1976**

The recommendation of the Ad Hoc Committee that an Advisory panel for Molecular Physics be appointed was not implemented. By default, the Heads of the Chemistry Department continued to function as de facto Directors of the Institute: Nathan L. Drake till 1959, G. Forrest Woods as Acting Chair and Director in 1959-1960, and Charles E. White till 1964. In 1964 the Institute finally became an independent unit with Homer W. Schamp as the first Director. At the same time a chemical physics graduate program was established administered by IMP in collaboration with the Chemistry and Physics Departments. However, shortly thereafter, Schamp left the University of Maryland at the College Park Campus to become Dean of Faculty at the newly established Campus of the University of Maryland in Baltimore County. Edward A. Mason served as Director till 1967, Joseph T. Vanderslice till 1968, Robert J. Munn till 1973 and William M. Benesch till 1976. The (full-time) faculty members of the Institute are listed in Table 2 together with their terms of service.

Table 2. Faculty at the Institute for Molecular Physics

| | |
|---|---|
| Homer W. Schamp, Jr. | 1952-1965 |
| Laurens Jansen | 1952-1958 |
| Edward A. Mason | 1955-1967 |
| Joseph T. Vanderslice | 1956-1968 |
| William G. Maisch | 1957-1963 |
| William M. Benesch | 1963-1976 |
| Lawrence C Krisher | 1963-1976 |
| Andrew G. De Rocco | 1963-1976 |
| Marshall L. Ginter | 1966-1976 |
| Ian L. Spain | 1966-1971 |
| William S. Benedict | 1967-1976 |
| Olav B. Verbeke | 1967-1969 |
| Robert J. Munn | 1967-1973 |
| Jan V. Sengers | 1968-1976 |
| Robert W. Gammon | 1971-1976 |
| Charles Maltz | 1971-1972 |
| Millard H. Alexander | 1971-1973 |

Laurens Jansen returned to Europe in 1958 [28]. It was Edward Mason who brought the Institute for Molecular Physics an international reputation in statistical molecular physics. He was a prolific scientist with an excellent physical intuition [29]. He also established a productive collaboration with Joseph T. Vanderslice, who became a faculty member in 1956, and with Robert J. Munn, who first appeared as a Harkness Fellow of the Commonwealth Fund on leave from the Department of Physical Chemistry at the University of Bristol in 1963-1964 and subsequently became a faculty member in 1967. In addition, Mason had an active collaboration with Louis Monchick of the Applied Physics Laboratory of the Johns Hopkins University in Laurel, MD. Ed Mason also attracted many visitors in statistical physics to the Institute.

Andrew G. De Rocco came in 1963 from the University of Michigan and worked in the area of macromolecular statistical physics. He was also the first faculty member embarking on biophysics research, such as membrane phase transitions, muscle contraction, noble-gas anesthesia, and circadian rhythms.

William G. Maisch, who had been a postdoc with Harry G. Drickamer, a pioneer in the field of pressure-tuning spectroscopy at the University of Illinois, was recruited in 1957, but he left in 1963, pursuing his further scientific career at the Naval Research laboratory. The subject of molecular spectroscopy got a major boost in the Institute for Molecular Physics after the arrival of William M. Benesch from the Johns Hopkins University and Lawrence C. Krisher from Harvard in 1963, followed by Marshall L. Ginter from the University of Chicago in 1966 and by William S. Benedict, a senior scientist who became a research professor in the Institute after his retirement from the Johns Hopkins University in 1967. This group made the Institute a prominent center in molecular spectroscopy.  One of the strength of the group is that they worked in close collaboration with complementary molecular-spectroscopy research at NASA in Greenbelt, MD, at the Naval Research Laboratory in Washington DC, and at the National Bureau of Standards in Gaithersburg, MD. As a result of the collaboration with NASA, Shelby G. Tilford from NASA was a part-time visiting professor in the Institute from 1969 till 1972. In 1974 Thomas J. McIlrath joined IFDAM, giving additional strength to molecular spectroscopy at the University of Maryland.

 Ian L. Spain was appointed as a faculty member in 1966 to continue the tradition in high-pressure research, a major goal for which the Institute was established originally. He brought a Dutch technician, Derk Kuil. However, he transferred to the Department of Chemical Engineering in 1971. Together with Jac Paauwe, an engineer who originally had worked at the 't Hart Company in Rotterdam on developing high-pressure equipment with Michels and who had become an engineer at NOL (later renamed Naval Surface Weapons Center), Spain produced a notable book on High-Pressure Technology [33,34]. Olav B. Verbeke was recruited from Belgium to enhance high-pressure research at the Institute also; however, he returned to Leuven after two years.

Ed Mason left the Institute in 1967 to accept a faculty position at Brown University in Providence, RI. He was succeeded by Joe Vanderslice as Director of the Institute. However, shortly thereafter Vanderslice became Head of the Department of Chemistry, while Munn succeeded him as Director of the Institute. Munn had become interested in computer-based education methods. He moved to the Department of Chemistry in 1973. I had the privilege of being appointed to the vacancy created by the departure of Mason and joined the Institute in January 1968. My wife, Anneke Levelt Sengers, and I had joined the Heat Division of NBS in December 1963 and I had become familiar with several faculty members at the University of Maryland including J. Robert Dorfman in IFDAM.  A memorable event was that we celebrated our first Thanksgiving in the US in the home of Homer and Julia Schamp. I did not continue in the field of high-pressure research, in which I had trained at the Van der Waals Laboratory as a graduate student. Instead I became interested in new opportunities for light-scattering experiments that had become available, applying Rayleigh scattering to study fluctuation phenomena in soft condensed matter.

Robert W. Gammon joined the Institute in 1971, having received his Ph.D. with Herman Z. Cummins at the Johns Hopkins University. Bob Gammon was an expert in Fabry-Perot interferometry. In addition his biggest project was concerned with critical-phenomena experiments in the space laboratory in collaboration with the NASA Lewis Research Center in Cleveland, OH.

Charles Maltz arrived in 1971, but left to become a medical doctor. Millard H. Alexander came in 1971 and transferred in 1973 to the Department of Chemistry, becoming a notable scientist in theoretical studies of molecular collisions

In 1976 IMP and IFDAM were merged to become the Institute for Physical Science and Technology (IPST). As a consequence, Bill Benesch, Andy De Rocco, Larry Krisher, Marshall Ginter, Bill Benedict, Jan Sengers, and Bob Gammon all became faculty members in IPST from 1976 onwards. Millard Alexander regained an affiliation with the Institute, but now with IPST in 2001. The new Institute also resurrected the Chemical Physics Program which had become dormant, this time including not only chemistry and physics but also engineering. Merging the resources and talents of the two institutes turned out to be a positive step towards further enhancement of the research stature of the University of Maryland in interdisciplinary science.

**Acknowledgments**

In preparing this account I have benefitted from valuable information received from many persons, including Anne J. Kox of the University of Amsterdam, Thijs Michels of the Technical University Eindhoven, Anneke Levelt Sengers of the National Institute of Standards and Technology, Homer W. Schamp Jr., Arnold E. Seigel, Jason G. Speck, Marshall and Dorothy Ginter of the University of Maryland. The historic file includes documents earlier received from C.A.M. Veraart-Michels, widow of Michels, concerning Michels' interactions with the University of Maryland, from P. Gopi Menon of Dow Chemical in Terneuzen and the University of Gent concerning some information about the discovery of polyethylene, and from Jac Paauwe of the Naval Surface Weapons Center concerning the interactions of Michels with the Company 't Hart in Rotterdam. I also received some comments from Roger P. and Barbara (Castle) Kohin in Worcester, MA, who were graduate students at the University of Maryland in the 1950's. I thank the members of the Communication Committee of IPST for their stimulating encouragement.

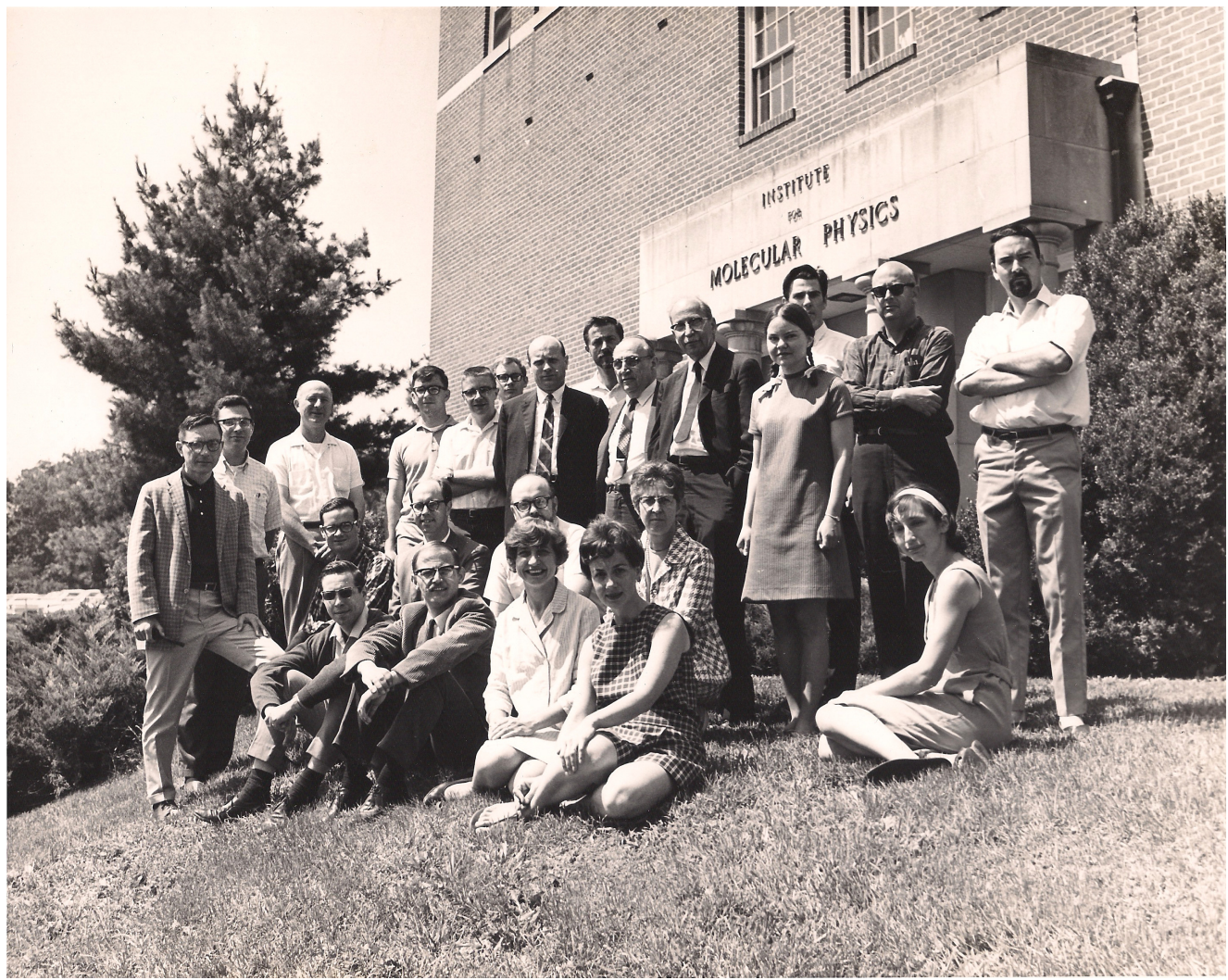

Figure 11. Staff of the Institute for Molecular Physics in 1968.

Form left to right sitting: #1 Dr. Olav B. Verbeke, #2 Derk Kuil (technician), #3 Mildred Pacl (accountant), #4 Nancy Brown (graduate student), #5 Dr. Ellen D. Yorke. From left to right kneeling on the grass: #1 R. Allan Kehs (graduate student), #2 Dr. Wilfred G. Norris, #3 Dr. Marshall L. Ginter, #4 Elisabeth Shell (secretary). From left to right standing: #1 Dr. Lawrence C. Krisher, #2 Russell Howard (graduate student), #3 John W. Trembly (glass blower since 1955), #4 Charles Brown (graduate student), #5 Stephen N. Wolf (graduate student), #6 Kenneth A. Saum (graduate student), #7 Dr. Jan V. Sengers, #8 Dr. Joseph T. Vanderslice, #9 Howard J. Kopp (technician since 1954), #10 Dr. William S. Benedict, #11 Dr. Ellen I. Saegebarth, #12 Dr. Ian L. Spain, #13 Dr. William M. Benesch, #14 Dr. Robert J. Munn (Dr. Andrew G. De Rocco missing). Explanatory notes: Dr. Wilfred G. Norris was a collaborator of Larry Krisher; Dr. Ellen Saegebarth was a protégé of Larry Krisher, both had received their Ph.D. with E. Bright Wilson at Harvard University; Dr. Ellen D. Yorke worked as a postdoc with Andy De Rocco.